\documentclass[acus]{pac99}
\usepackage{epsfig}
\newcommand{\ud}{\mathrm{d}}
\begin{document}
\title{NONLINEAR ACCELERATOR PROBLEMS VIA WAVELETS:\\
8. INVARIANT BASES, LOOPS AND KAM}
\author{A.~Fedorova,  M.~Zeitlin, IPME, RAS, St.~Petersburg, Russia
  \thanks{ e-mail: zeitlin@math.ipme.ru}
\thanks{http://www.ipme.ru/zeitlin.html;
http://www.ipme.nw.ru/zeitlin.html}}
\maketitle
\begin{abstract}
In this series of eight papers  we
present the applications of methods from
wavelet analysis to polynomial approximations for
a number of accelerator physics problems.
In this part we consider variational wavelet approach for loops,
invariant bases on semidirect product, KAM calculation via FWT.
\end{abstract}

\section{INTRODUCTION}
This is the eighth part of our eight presentations in which we consider
applications of methods from wavelet analysis to nonlinear accelerator
physics problems.
 This is a continuation of our results from [1]-[8],
in which we considered the applications of a number of analytical methods from
nonlinear (local) Fourier analysis, or wavelet analysis, to nonlinear
accelerator physics problems
 both general and with additional structures (Hamiltonian, symplectic
or quasicomplex), chaotic, quasiclassical, quantum. Wavelet analysis is
a relatively novel set of mathematical methods, which gives us a possibility
to work with well-localized bases in functional spaces and with the
general type of operators (differential, integral, pseudodifferential) in
such bases.
In contrast  with parts 1--4 in parts 5--8 we try to take into account
before using power analytical approaches underlying algebraical, geometrical,
topological structures related to kinematical, dynamical and hidden
symmetry of physical problems. In section 2 we consider wavelet approach
for calculation of Arnold--Weinstein curves (closed loops)
in Floer variational approach.
In section 3 we consider
the applications of orbit technique for constructing different types of invariant
wavelet bases in the particular case of affine Galilei group with
the semiproduct structure.
In section 4 we consider applications of very
useful fast wavelet transform (FWT) technique (part 6) to calculations in
KAM theory (symplectic scale of spaces).
This method gives maximally sparse representation of (differential) operator
 that allows us to take into account contribution from each level of
resolution.

\section{Floer Approach for Closed Loops}
Now we consider the generalization of  wavelet variational
approach  to the symplectic invariant calculation of
closed loops in Hamiltonian systems
[9]. As we demonstrated in [3]--[4] we have the parametrization
of our solution by some
reduced algebraical problem but in contrast to the cases from parts 1--4, where
the solution is parametrized by construction based on scalar
refinement equation, in symplectic case we have
parametrization of the solution
by matrix problems -- Quadratic Mirror Filters equations.
Now we consider a different approach.
Let$(M,\omega$) be a compact symplectic manifold of dimension $2n$, $\omega$ is
a closed 2-form (nondegenerate) on $M$ which induces an isomorphism $T^*M\to
TM$. Thus every smooth time-dependent Hamiltonian $H:{\bf R}\times M\to {\bf
R}$ corresponds to a time-dependent Hamiltonian vector field $X_H: {\bf
R}\times M\to TM$ defined by
$
\omega(X_H(t,x),\xi)=-{\rm}d_xH(t,x)\xi
$
 for
$\xi\in T_xM$. Let $H$ (and $X_H$) is periodic in time: $H(t+T,x)=H(t,x)$ and
consider corresponding Hamiltonian differential equation on $M$:
$
\dot x(t)=X_H(t,x(t))
$
The solutions $x(t)$ determine a 1-parameter
family of diffeomorphisms
$\psi_t\in {\rm Diff}(M)$ satisfying $\psi_t(x(0))=x(t)$. These diffeomorphisms are
symplectic:
 $\omega=\psi_t^*\omega$. Let $L=L_TM $ be the space of contractible loops in $M$
which are represented by smooth curves $\gamma: {\bf R}\to M$ satisfying
$\gamma(t+T)=\gamma(t)$. Then the contractible T-periodic solutions
can be characterized as the critical points of the functional
$S=S_T: L\to {\bf R}$:
\begin{equation}\label{eq:ST}
S_T(\gamma)=-\int_Du^*\omega+\int_0^TH(t,\gamma(t)){\rm d}t,\nonumber
\end{equation}
where $D\subset {\bf C}$ be a closed unit disc and $u: D\to M$ is a smooth
function, which on boundary agrees with $\gamma$, i.e. $u({\rm exp}\{2\pi i
\Theta\})=\gamma(\Theta T)$. Because  [$\omega$], the cohomology class of
$\omega$, vanishes then
$S_T(\gamma)$ is independent of choice of $u$.
Tangent space $T_\gamma L$ is the space of vector fields $\xi\in
C^\infty(\gamma^*TM)$ along $\gamma$ satisfying $\xi(t+T)=\xi(t)$.
Then we have for the 1-form ${\rm d}f: TL\to{\bf R}$
\begin{equation}\label{eq:dST}
{\rm d} S_T(\gamma)\xi=\int_0^T(\omega(\dot\gamma,\xi)+{\rm
d}H(t,\gamma)\xi){\rm d}t
\end{equation}
and the critical points of $S$ are contractible loops in $L$ which satisfy the
Hamiltonian equations. Thus the critical points are precisely
the required T-periodic solutions.
To describe the gradient of $S$ we choose $a$ on almost complex structure on $M$
which is compatible with $\omega$. This is an endomorphism $J\in C^\infty({\rm
End}(TM))$ satisfying $J^2=-I$ such that
$
g(\xi,\eta)=\omega(\xi,J(x)\eta),\  \xi,\eta\in T_xM
$
defines a Riemannian metric on M. The Hamiltonian vector field is then
represented by $X_H(t,x)=J(x)\nabla H(t,x)$, where $\nabla$
denotes the gradient  w.r.t. the x-variable using the metric.
Moreover the gradient of $S$ w.r.t. the induced metric on $L$ is given by
$
{\rm grad} S(\gamma)=J(\gamma)\dot\gamma+\nabla H(t,\gamma),\
\gamma\in L
$.
Studying the critical points of $S$  is confronted with the well-known
difficulty that the variational integral is neither bounded from below nor from
above. Moreover, at every possible critical point the Hessian of $f$ has an
infinite dimensional positive and an infinite dimensional negative subspaces, so
the standard Morse theory is not applicable.
The additional problem is that the gradient vector field on the loop space $L$:
${\rm d}\gamma/{\rm d}s=-{\rm grad}f(\gamma)$
does not define a well posed Cauchy problem.
But Floer [9] found a way to analyse the space ${\mathcal M}$ of bounded solutions
consisting of the critical points together with their connecting orbits.
He used a combination of variational approach and Gromov's elliptic technique.
A gradient flow line of $f$ is a smooth solution $u: {\bf R}\to M$ of the
partial differential equation
\begin{equation}\label{eq:duds}
\frac{\partial u}{\partial s}+J(u)\frac{\partial u}{\partial
t}+\nabla H(t,u)=0,
\end{equation}
which satisfies $u(s,t+T)=u(s,t)$. The key point is to consider (\ref{eq:duds})
not as the flow on the loop space but as an elliptic boundary value problem. It
should be noted that (\ref{eq:duds}) is a generalization of equation for
Gromov's pseudoholomorphic curves (correspond to the case $\nabla
H=0$ in (\ref{eq:duds})).
Let ${\mathcal M}_T={\mathcal M}_T(H,J)$ the space of bounded solutions of (\ref{eq:duds}), i.e. the
space of smooth functions $u: {\bf C}/ iT{\bf Z}\to M$, which are contractible,
solve equation (\ref{eq:duds}) and have finite energy flow:
\begin{equation}\label{eq:PhiT}
\Phi_T(u)=\frac{1}{2}\int\int_0^T\Big(\arrowvert\frac{\partial u}
{\partial
 s}\arrowvert^2+\arrowvert\frac{\partial u}
{\partial t}-X_H(t,u)\arrowvert^2\Big){\rm d}t{\rm
 d}s.
\end{equation}
For every $u\in M_T$ there exists a pair $x,y$ of contractible T-periodic
solutions, such that $u$ is a connecting orbit from $y$ to
$x$:
$
\lim_{s\to-\infty}u(s,t)=y(t), \  \lim_{s\to+\infty}=x(t)
$.
Then our  approach from preceding parts, which we may apply or on the level of standard
boundary problem  or on the level of variational approach and
 representation of operators (in our case, $J$ and $\nabla$)
 according to part 6(FWT technique) lead
us to wavelet representation of closed loops.

\section{Continuous Wavelet Transform. Bases for Solutions.}
  When we  take into account the Hamiltonian
or Lagrangian structures from part 7
 we need to consider generalized wavelets, which
allow us to consider the corresponding  structures  instead of
compactly supported wavelet representation from parts 1--4.
We consider an important particular case of constructions
from part 7: affine
relativity group (relativity group
combined with dilations) --- affine Galilei group in n-dimensions. So, we have
combination of Galilei group with independent space and time dilations:
$G_{aff}=G_m\bowtie D_2$,
where $D_2=({\bf R}^{+}_*)^2\simeq {\bf R}^2$, $G_m$ is extended
Galilei group corresponding to mass parameter $m>0$ ($G_{aff}$ is noncentral
extension of $G\bowtie D_2$ by ${\bf R}$, where G is usual Galilei group).
Generic element of $G_{aff}$ is $g=(\Phi,b_0,b;v;R,a_0,a)$, where
$\Phi\in{\bf R}$ is the extension parameter in $G_m$, $b_0\in{\bf R}$,
$b\in{\bf R}^n$ are the time and space translations, $v\in{\bf R}^n$ is the boost
parameter, $R\in SO(n)$ is a rotation and $a_0,a\in{\bf R}^+_*$ are time and
space dilations. The actions of $g$ on space-time  is then $x\mapsto
aRx+a_0vt+b$, $t\mapsto a_0t+b_0$, where $x=(x_1,x_2,...,x_n)$.
It should be noted that $D_2$ acts nontrivially on $G_m$.
Space-time wavelets associated to $G_{aff}$ corresponds to unitary irreducible
representation of spin zero. It may be obtained via orbit method. The Hilbert
space is ${\mathcal H}=L^2 ({\bf R}^n\times{\bf R},
{\rm d}k{\rm d}\omega)$, $k=(k_1,...,k_n)$, where ${\bf R}^n\times{\bf R}$ may
be identified with usual Minkowski space and we have for representation:
\begin{equation}
(U(g)\Psi)(k,\omega)=\sqrt{a_0a^n}{\rm exp}i(m\Phi+kb-\omega
b_0)\Psi(k',\omega'),
\end{equation}
with  $k'=aR^{-1}(k+mv)$, $\omega'=a_0(\omega-kv-\frac{1}{2}mv^2)$,
$m'=(a^2/a_0) m$.
Mass m is a coordinate in the dual of the Lie algebra and these relations are a
part of coadjoint action  of $G_{aff}$. This representation is unitary and
irreducible but not square integrable. So, we need to consider reduction to the
corresponding quotients $X=G/H$. We consider the case in which H=\{phase
changes $\Phi$ and space dilations $a$\}. Then the space $X=G/H$ is parametrized
by points $\bar{x}=(b_0,b;v;R;a_0)$. There is a dense set of vectors
$\eta\in{\mathcal H}$ admissible ${\rm mod}(H,\sigma_\beta)$, where
$\sigma_\beta$ is the
corresponding section.
We have a two-parameter family of functions $\beta$(dilations):
$\beta(\bar{x})=(\mu_0+\lambda_)a_0)^{1/2}$, $\lambda_0, \mu_0\in{\bf R}$.
Then any admissible vector $\eta$ generates a tight frame of Galilean wavelets
\begin{equation}
\eta_{\beta(\bar{x})}(k,\omega)=\sqrt{a_0(\mu_0+\lambda_0a_0)^{n/2}}
{\rm e}^{i(kb-\omega b_0)}\eta(k',\omega'),
\end{equation}
with $k'=(\mu_0+\lambda_0 a)^{1/2}R^{-1}(k+mv)$,
$\omega'=a_0(\omega-kv-mv^2/2)$.
The simplest examples of admissible vectors (corresponding to usual Galilei
case) are Gaussian vector: $\eta(k)\sim{\rm exp}(-k^2/2mu)$ and
binomial vector: $\eta(k)\sim(1+k^2/2mu)^{-\alpha/2}$, $\alpha> 1/2$, where $u$
is a kind of internal energy.
When we impose the relation $a_0=a^2$ then we have the restriction to the
Galilei-Schr\"odinger group $G_s=G_m\bowtie D_s$, where $D_s$ is the
one-dimensional subgroup of $D_2$. $G_s$ is a natural invariance group of
both the Schr\"odinger equation and the heat equation.
The restriction to $G_s$ of the representation (29) splits into the direct
sum of two irreducible ones $U=U_+\oplus U_-$ corresponding to the
decomposition $L^2({\bf R}^n\times{\bf R}, {\rm d}k{\rm d}\omega)=
{\mathcal H}_+\oplus{\mathcal H}_-$, where
$
{\mathcal H}_\pm=L^2(D_{\pm}, {\rm d}k{\rm d}\omega\
=\{ \psi\in L^2({\bf R}^n\times{\bf R},{\rm d}k{\rm d}\omega),\
\psi(k,\omega)=0\  {\textrm for} \  \omega+k^2/2m=0\}.
$
These two subspaces are the analogues of usual Hardy spaces on
${\bf R}$, i.e. the subspaces of (anti)progressive wavelets (see also below,
part III A).
The two representation $U_\pm$ are square integrable modulo the center.
There is a dense set of admissible vectors $\eta$, and each of them generates a
set of $CS$ of Gilmore-Perelomov type.
Typical wavelets of this kind are:\ the Schr\"odinger-Marr wavelet:
$
\eta(x,t)=(i\partial_t+{\triangle}/{2m}){\rm e}^{-(x^2+t^2)/2}
$,
the Schr\"odinger-Cauchy wavelet:
$
\psi(x,t)=(i\partial_t+{\triangle}/{2m})\times
{(t+i)\prod_{j=1}^n(x_j+i)}^{-1}
$.
So, in the same way we can construct different invariant bases with explicit
manifestation of underlying symmetry
for solving Hamiltonian  or Lagrangian equations.

\section{SYMPLECTIC HILBERT SCALES VIA WA\-VE\-LETS}

We can solve many important dynamical problems such that KAM
perturbations, spread of energy to higher modes, weak turbulence, growths of
solutions of Hamiltonian equations only if we consider scales of spaces instead
of one functional space. For Hamiltonian system and their perturbations
for which we need take into account underlying symplectic structure we
need to consider symplectic scales of spaces. So, if
$\dot{u}(t)=J\nabla K(u(t))$
is Hamiltonian equation we need wavelet description of symplectic or
quasicomplex structure on the level of functional spaces. It is very
important that according to [12] Hilbert basis is in the same time a
Darboux basis to corresponding symplectic structure.
We need to provide Hilbert scale $\{Z_s\}$ with symplectic structure [12].
All what we need is the following.
 $J$  is a linear operator, $J : Z_{\infty}\to Z_\infty$,
$J(Z_\infty)=Z_\infty$, where $Z_\infty =\cap Z_s$.
$J$ determines an isomorphism of  scale $\{Z_s\}$ of order $d_J\geq 0$.
 The operator $J$ with domain of definition $Z_\infty$ is
antisymmetric in $Z$:
$
<J z_1,z_2>_Z=-<z_1,J z_2>_Z, z_1,z_2 \in $ $ Z_\infty
$.
Then the triple
$\{Z,\{Z_s|s\in R\},\
\alpha=<\bar J dz,dz>\}
$
 is symplectic Hilbert scale. So, we may consider
any dynamical Hamiltonian problem on functional level.
As an example, for KdV equation we have
$ Z_s=\{u(x)\in H^s(T^1)|\int^{2\pi}_0 u(x)\ud x=0\},\
s\in R,$
$ J=\partial/\partial x,$
 is isomorphism  of the scale of  order one, $\bar J=-(J)^{-1}$ is
isomorphism of order $-1$.
According to [13] general functional spaces and scales of spaces such as
Holder--Zygmund, Triebel--Lizorkin and Sobolev can be characterized
through wavelet coefficients or wavelet transforms. As a rule, the faster
the wavelet coefficients decay, the more the analyzed function is
regular [13]. Most important for us example is the scale of Sobolev spaces.
Let $H_k(R^n)$ is the Hilbert space of all distributions with finite norm
$
\Vert s\Vert^2_{H_k(R^n)}=\int \ud\xi(1+\vert\xi\vert^2)^{k/2}\vert
\hat s(\xi)\vert^2.
$
Let us consider wavelet transform
$$
W_g f(b,a)=\int_{R^n}\ud x\frac{1}{a^n}\bar g\left(\frac{x-b}{a}\right) f(x),
$$
$ b\in R^n, \quad a>0$,
w.r.t. analyzing wavelet $g$, which is strictly admissible, i.e.
$
C_{g,g}=\int_0^\infty({\ud a}/{a})\vert\bar{\hat g(ak)}\vert^2<\infty.
$
Then there is a $c\geq 1$ such that
\begin{eqnarray*}
&&c^{-1}\Vert s \Vert^2_{H_k(R^n)}\leq\int_{H^n}
\frac{\ud b\ud a}{a}(1+a^{-2\gamma})\vert\times\\
&& W_gs(b,a)\vert^2
\leq
c\|s\|^2_{H_k(R^n)}.
\end{eqnarray*}
This shows that localization of the wavelet coefficients at small
scale is linked to local regularity.
So, we need representation for differential operator ($J$ in our case) in
wavelet basis. We consider it by  means of the methods from part 6.

We are very grateful to M.~Cornacchia (SLAC),
W.~Her\-r\-man\-nsfeldt (SLAC)
Mrs. J.~Kono (LBL) and
M.~Laraneta (UCLA) for
 their permanent encouragement.

\end{document}